\documentclass[prl,twocolumn]{revtex4}

\usepackage{amsmath}
\usepackage{amstext}
\usepackage{latexsym}
\usepackage[dvips]{graphicx}

\newcommand{\beq}{\begin{equation}}
\newcommand{\eeq}{\end{equation}}
\newcommand{\bea}{\begin{eqnarray}}
\newcommand{\eea}{\end{eqnarray}}

\begin{document}
\title{Optimal control technique for Many Body Quantum Systems dynamics}

\author{Patrick Doria}
\affiliation {Institut f\"ur Quanteninformationsverarbeitung
Albert-Einstein-Allee 11
D-89069 Ulm, Germany.}
\affiliation{Politecnico di Torino, Corso Duca degli Abruzzi, 24
10129 Torino, Italy.}

\author{Tommaso Calarco}
\affiliation {Institut f\"ur Quanteninformationsverarbeitung
Albert-Einstein-Allee 11
D-89069 Ulm, Germany.}

\author{Simone Montangero}
\affiliation {Institut f\"ur Quanteninformationsverarbeitung
Albert-Einstein-Allee 11
D-89069 Ulm, Germany.}
\email{simone.montangero@uni-ulm.de}

\begin{abstract}  %(approx 150 words)
We present an efficient strategy for controlling a vast range of non-integrable 
quantum many-body one-dimensional systems that can
be merged with state-of-the-art tensor network simulation  
methods like the Density Matrix Renormalization Group. 
To demonstrate its potential, we employ it to solve a major issue in
current optical-lattice physics with ultra-cold atoms: we show how to 
reduce by about two orders of magnitudes the time needed to bring a 
superfluid gas into a Mott insulator state, while suppressing defects 
by more than one order of magnitude as compared 
to current experiments~\cite{MSexp}. Finally, we show that the optimal
pulse is robust against atom number fluctuations.
\end{abstract} 

\maketitle

Classical control theory has played a major role in the development of
present-day technologies~\cite{ccontrol}. Likewise, recently developed
quantum optimal control methods~\cite{dalessandro,krotov, rabitz} can be
applied to emerging quantum technologies, e.g. quantum information
processing -- until now, at the level of a few qubits
\cite{NMR,QCapplyed1, QCapplyed2}.   
However, such methods encounter severe limits when applied to
many-body quantum systems: due to the complexity of simulating the
latter, existing quantum control algorithms (requiring many iterations
to converge) usually fail to yield a desired final state within an
acceptable computational time. A paradigmatic application of control
of many body quantum system is the control of the 
dynamics of a quantum phase transition. 
%We approach this problem merging
%tensor methods and an ad-hoc developed optimization
%strategy, and we apply it to the study of quantum phase transition
%dynamics in cold atoms in optical lattices.
The process of crossing a phase transition in an optimal way has
been studied for decades for classical systems.
Only recently it has been recast in the quantum domain, 
attracting a lot of attention (see e.g.~\cite{kzmech} and
references therein) since, for instance, it has implications
for adiabatic quantum computation and quantum annealing~\cite{farhi}.
A transition between different phases is usually performed 
by ``slowly'' (adiabatically) sweeping an external control 
parameter across the critical point, allowing for a
transformation from the initial to the final system ground state with 
sufficiently high probability. However, at the critical point in the 
thermodynamical limit a perfect adiabatic process is forbidden in 
finite time~\cite{adtheo}. 
Thus, the resulting final state (for finite-time transformations) 
is characterized by some residual excitation energy,
corresponding to the formation of topological defects within finite-size
domains. The Kibble-Zurek theory has been shown to yield good
estimates of the density of defects or of the residual 
energy~\cite{revzurek,kzmech}.
The importance of these estimates in the quantum domain 
is underscored by the fact that, apart from very specific cases where 
analytical solutions are available, theoretical investigations must 
rely on heavy numerical simulations due to the exponential 
growth of the Hilbert space with the system size 
and to the diverging entanglement at the critical 
point~\cite{vidal}. Nevertheless, it is possible to
perform one-dimensional simulations of the dynamics by means of 
tensor-network-based techniques such as the time-dependent Density
Matrix Renormalization Group (t-DMRG)~\cite{revscholl}. 

%{Moreover, it introduces an optimization strategy that can be
%integrated straightforwardly into an experiment to implement closed
%loop quantum control, thereby taking automatically into account
%unknown systematic errors and further robustly enhancing the
%experimental results.}

The basic underlying idea of classical control is to pick a 
specific path in parameter space to perform a specific task. 
This is formally  a cost functional extremization that depends on the state 
of the system and is attained by varying some external control parameters.
In a quantum-mechanical context, a big advantage is that the goal can 
be reached via interference of many different paths in parameter
space, rather than just one. In few-body quantum systems, it has 
been shown that optimal control finds 
optimal paths in the parameter space that result in 
constructive interference of the system's classical trajectories 
toward a given goal \cite{krotov}. 
Indeed, present optimal control strategies demonstrated an impressive
control of quantum systems, ranging from optimization of NMR
pulses \cite{NMR} to atomic \cite{MarkerAtoms} and superconducting
qubits \cite{QCapplyed1}, 
as well as the crossing of a quantum phase transition (QPT) in the analytically 
solvable quantum Ising model \cite{caneva}.
However, despite their effectiveness, they cannot be efficiently applied to 
systems that require tensor network methods for their simulation. 

The present letter marks a step further 
--- it provides for the first time a means to control the evolution of 
a non-integrable many-body quantum system, resulting in the
optimization of a given figure of merit. % in such a way to reach a
%desired final state % across a quantum phase transition (QPT)  
%with very high probability over a finite time. 
This is done by introducing a strategy to integrate
optimal control with t-DMRG simulations of the many-body quantum
dynamics. 

{\it Tensor-network-simulations}-- Tensor network methods are 
based on the assumption that it is possible to
describe approximately a wide class of states with a simple 
tensor structure. In particular the DMRG describes  ground states static
properties of one dimensional
systems by means of a Matrix Product State (MPS)~\cite{MPS}. 
The main characteristic of an MPS is that 
the resources needed to describe 
a given system depend only polynomially on the system size $N$,
due to the introduction of an ancillary dimension $m$ that
determines the precision of the approximation.
Since an exact description requires exponentially increasing resources
with the number of components $N$, the tensor network approach results
in  an exponential gain in resources. Given a system Hamiltonian, 
the best possible approximated description of the system ground 
state --within the MPS at fixed $m$-- is determined by means of an efficient 
energy minimization. With some slight modification, { discretizing 
the time $T= n_{steps} \Delta t$ and performing a Trotter expansion,} 
the algorithm can be adapted to follow a 
state time evolution, the so-called t-DMRG~\cite{revscholl}. 
The t-DMRG is a very powerful numerical method for
efficiently numerically simulating the time evolution of 
one-dimensional many body quantum systems. %composed by $N$ interacting 
%local systems or sites. 
The class of states and of time evolutions that can be
efficiently described with a small error is determined by the
presence of entanglement between the different system components
\cite{vidal}. Here, we will use the t-DMRG for the simulation of 
cold atoms in time dependent optical lattices, which we feed into
the Chopped RAndom Basis (CRAB) optimization algorithm as described below.

%%%%%%%%%%%%%%%%%%%%%%%%%%%%%
\begin{figure}[t]
    \begin{center}
    \includegraphics[scale=0.8]{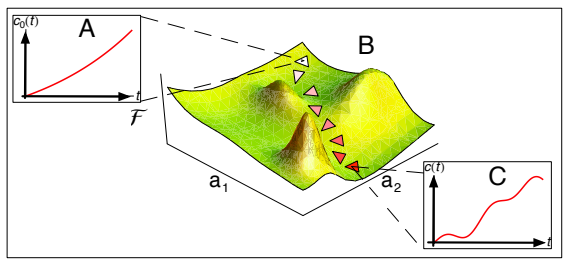}
    \caption{A) An initial guess pulse $c^0(t)$ is used as a starting
    point. B) The function
    $\mathcal{F}(\vec a)$ for the case $\vec a =\{a_1, a_2\}$ and the initial
    polytope (white triangle) are defined and moved ``downhill''
    (darker red triangles) until
    convergence is reached. C) The final point is recast
    as the optimal pulse $c(t)$. % and applied to the physical system.
}
        \label{CRAB}
    \end{center}
\end{figure}
%%%%%%%%%%%%%%%%%%%%%%%%%%%%%
%
%Here we present an optimal search in a truncated dual space, the 
%Chopped RAndom Basis (CRAB) optimization, that can be efficiently
%applied to t-DMRG simulations. 
{\it The CRAB method}-- The general scenario of an optimal control problem can be stated as
follows: given a system described by a Hamiltonian $H$ depending on some 
control parameters $c_j(t)$ with $j=1,\dots, N_C$, 
the goal is to find the $c_j$'s time dependence (pulse shape) that
extremizes a given figure of merit $\mathcal{F}$, 
for instance the final system energy, state fidelity, or
entanglement. We then start with an initial 
pulse guess $c_j^0(t)$ and look for the best correction that has a
simple expression in a given functional basis. As an explicative
example, here we focus on the case where the correction is of the form
%\begin{equation}
$c_j(t)= c_j^0(t) \cdot f_j(t)$,
%\label{corr}
%\end{equation}
and the functions $f_j(t)$ can be simply expressed in a truncated
Fourier space, depending on the expansion coefficients 
$\vec a_j = a_j^k$ ($k=1, \dots, M_j$). 
In particular, in the following, we start from an initial ansatz, 
e.g. an exponential or linear ramp, and 
we introduce a correction of the form%~(\ref{corr}) where
\begin{equation}
f(t)= 
\frac{1}{\mathcal{N}}
\left[1+\sum_k A_k \sin(\nu_k t)+ B_k \cos(\nu_k t)\right]. 
\label{expansion}
\end{equation}
%%%%%%%%%%%%%%%%%%%%%%%%%%%%%
\begin{figure}[t]
    \begin{center}
    \includegraphics[scale=0.7]{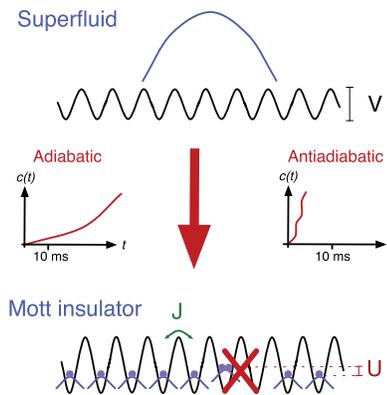}
    \caption{Scheme of the Mott-superfluid transition in the
      homogeneous system for average
      occupation number $\langle n \rangle =1$: increasing the
      lattice depth $V$ (black line) the atoms superfluid wave
      functions (upper) localize in the wells (lower). If the
      transition is not adiabatic --or optimized-- defects appear (here represented by a
      hole and a doubly occupied site).}
        \label{SMtrans}
    \end{center}
\end{figure}
%%%%%%%%%%%%%%%%%%%%%%%%%%%%%
%

\noindent Here, $k=1, \dots, M $,  $\nu_k = 2 \pi k (1+r_k)/T $ are 
``randomized'' Fourier harmonics, $T$ is the total time evolution, $r_k \in [0:1]$ are 
random numbers  with a flat distribution, and $\mathcal{N}$ is a
normalization constant to keep the initial and final control pulse
values fixed. The optimization problem is then reformulated as the
extremization of a multivariable function $\mathcal{F}(\{A_k\},\{B_k\},\{\nu_k)\}$, 
which can be numerically approached with a suitable
method, e.g., steepest descent or conjugate gradient~\cite{numrec}. 
When using CRAB together with t-DMRG, computing
the gradient of $\mathcal{F}$ is extremely resource consuming, if not
impossible. Thus we resort to a direct search method like the
Nelder-Mead or Simplex methods~\cite{numrec}. 
They are based on the construction of a
polytope defined by some initial set of points in the space of
parameters that ``rolls down the hill'' following 
predefined rules until reaching a (possibly local) minimum (see
Fig.~\ref{CRAB}). Due to the fact that direct search
methods are based on many {\it independent} evaluations of the function
to be minimized, they can be efficiently implemented together with
t-DMRG simulations (and possibly performed in parallel). 
We stress that the functional dependency 
of the correction presented here (Eq.~(\ref{expansion}))
is one possible approach: different strategies might be explored. Indeed,
making a given choice confines the search of the optimal driving field
in a subspace of the whole space of functions
and the results might depend on this
choice. On the other hand, this approach simplifies the optimization 
problem that would be otherwise computationally unfeasible when 
t-DMRG simulations are needed. As shown below, the 
described choice allows to perform a successful optimization.

{\it The optical-lattice system}-- Very recently, the experimental and theoretical analysis of the
dynamics of cold atoms in optical lattices has experienced a 
fast development, after the experimental demonstration 
of coherent manipulation of ultra-cold atoms in the
seminal work of Ref.~\cite{greiner}, where a Bose-Einstein 
condensate is first loaded into a single trap, and then a 
periodic lattice potential is slowly ramped up, 
inducing a quantum phase transition to a Mott insulator. This is the
enabling step for a wide range of experiments, from
transport or spectroscopy to quantum information
processing~\cite{blochrev}. 
In most of these applications, it is essential to achieve the lowest possible
number of defects in the final state, that is, to reach exactly a
final state with fixed number of atoms per site, e.g. unit filling. 
Up to now, this has been pursued by limiting the process speed -- 
the superfluid-Mott insulator
transition has been performed in about a hundred ms, with a density
of defects typically of the order of 10\%~\cite{toolbox}. 

%
%%%%%%%%%%%%%%%%%%%%%%%%%%%%%
\begin{figure}[t]
    \begin{center}
    \includegraphics[scale=0.28]{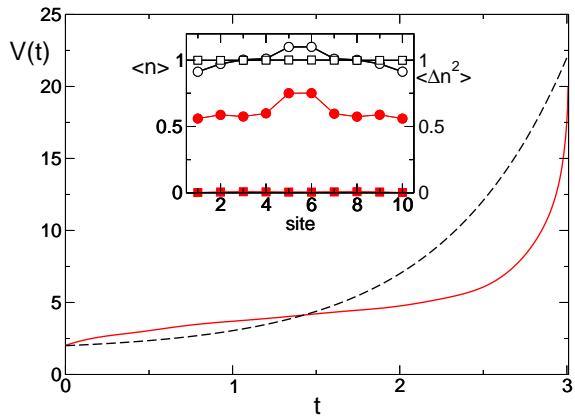}
    \caption{{ Initial guess
        (dashed black) and optimal ramp  (solid red) $V(t)$ for the Bose Hubbard model in the
      presence of the trap % (experimental  parameters from \cite{MSexp})
      with $N=30$ sites, total time evolution $T \simeq 3\,
      \mathrm{ms}$.} Inset: Populations $\langle n_i \rangle$ (empty black) 
     and fluctuations $\langle \Delta n_i^2 \rangle$ (full
      red) at time $t=T$ for the exponential initial guess (circles) and
      optimal ramp (squares) for $N=10$.}
        \label{pulse}
    \end{center}
\end{figure}
%%%%%%%%%%%%%%%%%%%%%%%%%%%%%
%
Cold atoms in an optical lattice can be described by the Bose Hubbard model
defined by the Hamiltonian~\cite{Jaksch98, blochrev}
\begin{equation}
\mathrm{H}\!=\!\sum_{j}\!\left[
-J(b_{j}^{\dagger}b_{j+1}\!+\!\mathrm{h.c.})
%b_{j}b^{\dagger}_{j+1}
\!+\Omega (j-\frac{N}{2})^2
n_j %\nonumber \\ 
\!+\!\frac{U}{2}(n_j^2-n_j)\!\right]\!\!.
\label{HamBHInt}
\end{equation}
The first term on the r.h.s. of Eq.(\ref{HamBHInt}) describes the
tunneling of bosons between neighbouring sites with rate $J$; $\Omega$ 
is the curvature of the external trapping potential, and $n_{j} = b^{\dag}_{j}
b_{j}$ is the density operator with bosonic creation (annihilation) 
operators $b^{\dag}_{j}$ ($b_{j}$) at site $j=1, \dots, N$. The last term is the
on-site contact interaction with energy $U$. The system parameters $U$
and $J$ can be expressed as a function of the optical lattice depth
$V$ (we set $\hbar=1$ from now on)~\cite{blochrev}. As sketched in
Fig.~\ref{SMtrans}, the system undergoes a quantum phase transition
from a superfluid phase to a Mott insulator as a function of the ratio 
$J/U$. In a homogeneous one-dimensional system, the QPT 
is expected to occur at $J_c/U \simeq 0.083 $, where (upon decreasing the
ratio $J/U$) the ground state wave function drastically changes from a 
Fermi-Thomas distribution with high fluctuations in the number 
of particles per site to a simple product of local Fock states with no
fluctuations in the number of atoms per site~\cite{blochrev}.
In the presence of an external trapping potential on top 
of the optical lattice, the phase diagram is more complex:
the two phases coexists in different trap regions
and typical ``cake'' structures are formed~\cite{batrouni}.

{\it Results}-- Following previous numerical studies~\cite{kollath} that modeled the
experiment~\cite{MSexp}, and supported by strong evidence of agreement
between numerical simulations and experimental results~\cite{fertig05,
clark}, we studied both the ideal homogeneous system ($\Omega=0$) and 
the experimental setup of~\cite{fertig05} where the trapping potential
is present. We applied the CRAB optimization to the preparation of a 
Mott insulator with ultra-cold atoms in an optical
lattice, that is, we optimized the ratio $J/U(t)$ that drives 
the superfluid-Mott insulator transition. The resulting optimal ramp shape 
drives the system into a final Mott insulator state with a 
density of defects below half a percent in a total time of the 
order of a few milliseconds, amounting to a drastic improvement 
in the process time and in the quality of the final state -- by 
about two orders of magnitude and by more than one, respectively.

%%%%%%%%%%%%%%%%%%%%%%%%%%%%
\begin{figure}[t]
    \begin{center}
    \includegraphics[scale=.23]{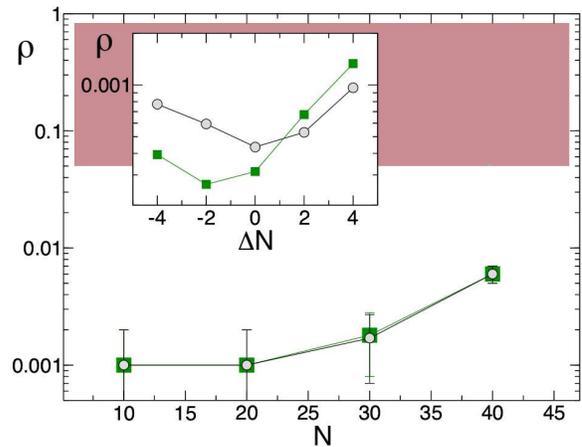}
    \caption{Residual defect density $\rho$  for %as a function of %the system size
    $N=10,20,30,40$, $T \simeq 3 ms$, $\rho_c = 0.001$ 
    for the homogeneous system 
    (green squares) and in the presence of the trap (grey circles). 
%    with experimental  parameters from~\cite{MSexp}. 
    The red region highlights the typical $\rho$ %unoptimized density of defects 
    for different initial ramp shapes (see text). 
    Inset: Final $\rho$ computed applying the pulse optimized for system size $N=20$ 
    to different system sizes $\Delta N= -4, \dots, 4$ (at
    constant filling). %All errors are  mainly due to the Trotter expansion: 
    The results are almost independent from the truncation error for $m>50$.
}
    \label{Defectdensity}
    \end{center}
\end{figure}
%%%%%%%%%%%%%%%%%%%%%%%%%%%%%
%
%Starting from the Hamiltonian~(\ref{HamBHInt}) for the homogeneous case, we decrease exponentially  the ratio $J/U(t)$ with $J/U(0)=2$ ---
%an initial condition very 
%far from the phase transition and deep in the superfluid region. In the case of 
%the non-homogeneous system, 

%\begin{center}
%{\bf Results and Discussion}
%\end{center}

%
We consider a starting value of the lattice depth $V(0)=2 E_r$ 
corresponding to $J/U(0)\sim 0.52$, since the description of the
experimental system by (Eq.~\ref{HamBHInt}) breaks down for $V(0) \lesssim 2
E_r$ \cite{Jaksch98}. However, the initial lattice switching on ($V=0
\to 2E_r$) can be performed very quickly without exciting the system 
(few milliseconds at most)~\cite{private}. 
We optimize the ramp to obtain the minimal residual 
energy per site $\Delta E/N = (E(T)-E_G)/N$ (where $E_G$ is the exact
final ground state energy). {In all simulations performed} we set the total time $T= 50 \hbar/U
\simeq 3.01\,ms $ and the final lattice depth 
$V(T)/ E_r= 22 \sim 2.4 \cdot 10^{-3} J/U$, deep inside the Mott
insulator phase. Unless explicitly stated, we set the average
occupation to one ($\sum_i \langle n_i \rangle = N$).
In all DMRG simulations, we exploited the conservation of number of
particles and used $m=20, \dots, 100$, $\Delta t= 10^{-2}
\div 10^{-3}$. 
We computed the final density of defects 
$\rho = \frac{1}{N}\sum_i |\langle n_i \rangle -1|$: 
when it reached a given threshold $\rho_c=10^{-3}$, the 
optimization was halted. 
In Fig.~\ref{pulse} we report a typical result of the optimization
process: the initial guess and final optimal ramp for the 
system in the presence of the confining trap are shown
for the parameter values corresponding to the experiment \cite{MSexp},
for a system size $N=30$. As it can be clearly seen, the pulse is 
modulated with respect to the initial exponential guess and no high
frequencies are present, reflecting the constraint introduced by the
CRAB optimization. 
In the inset we display the final occupation
numbers and the corresponding fluctuations, for the initial
exponential ramp and the optimal pulse in the case $N=10$.
The figure clearly demonstrates the
convergence to a Mott insulator in the latter case
as fluctuations are drastically reduced and the occupation is
exactly one for every site. 

Finally, in Fig.~\ref{Defectdensity} we plot the optimized density of 
defects $\rho$ as a function of the system size (up to $N=40$)
for the homogeneous and for the trapped system, demonstrating an 
improvement with respect to the initial guess 
by one (two) orders of magnitudes. 
Indeed, the exponential guess --~like other guesses: linear,
random, { and a pulse optimized for a smaller system ($N=8$ sites)}~-- 
gave residual density of defects of the order of $10\%$
(red region in Fig.~\ref{Defectdensity}). 
{ To check the experimental feasibility of our findings, 
  we studied the stability of the optimal evolutions under
  different sources of error and experimental uncertainties, like atom
  number fluctuations. The inset of Fig.~\ref{Defectdensity} shows the
  final density of defects when an optimal pulse computed for a given
  system size, is applied to a different system size (keeping the
  average filling constant). As it can be seen, the optimization works
  also for system size fluctuations of up to $20 \%$: 
  the final density of defects is of the same order. This robustness is
  crucial as the experimental realization of these systems is 
  performed in parallel on many different one-dimensional tubes with different
  numbers of atoms~\cite{blochrev}. We also checked the cases
  of different filling and of pulse distortion obtaining similar results (data not shown).}

{\it Outlook}-- In conclusion, we would like to mention that the CRAB optimization
strategy introduced here can in principle be applied also to open 
quantum many-body systems, e.g. by means of recently introduced
numerical techniques \cite{vidal1}. Perhaps an even more stimulating 
perspective would be that of implementing it with a quantum system in place of
the t-DMRG classical simulator, i.e. performing a CRAB based
closed-loop optimization~\cite{Brif2010}. 
{The optimization might be performed during the experimental
  repetitions of the measurement
 processes, thus adding a small overhead to the experimental complexity}.  
This would extend the applicability of the CRAB method to the 
optimization of quantum phenomena that are completely out of reach for simulation on classical computers, 
and represent a major design tool for future quantum technologies.

%\begin{center}
%{\bf Methods}
%\end{center}

%{\bf Acknowledgements}
We thank F.~Dalfovo, J.~Denschlag, R.~Fazio, C.~Fort, M.~Greiner, C.~Koch,
C.~Menotti, G.~Pupillo and M.~Rizzi for discussions; 
the PwP project (www.dmrg.it), the DFG
(SFB/TRR~21), the bwGRiD, the EC (grant 247687, AQUTE, and PICC) for support.


\begin{thebibliography}{99}


\bibitem{MSexp} T.~St\"oferle, {\it et. al.}, % Henning Moritz, Christian Schori, Michael K\"ohl, and Tilman Esslinger
%``Transition from a Strongly Interacting 1D Superfluid to a Mott Insulator''
Phys. Rev. Lett. {\bf 92}, 130403 (2004).

\bibitem{ccontrol}  J.T.~Betts ``Practical Methods for Optimal Control using Nonlinear Programming'',  
SIAM, Philadelphia, (2001).

\bibitem{dalessandro} D.~D'Alessandro 
``Introduction to Quantum Control and Dynamics'', 
Chapman \& Hall/CRC (2007).
       
\bibitem{rabitz} C.~Brif, R.~Chakrabarti, and H.~Rabitz 
%``Control of quantum phenomena: Past, present, and future'' 
arXive:0912.5121.

\bibitem{krotov} V. F. Krotov, ``Global Methods in Optimal Control
  Theory'' (Marcel Dekker, New York, 1996).

\bibitem{QCapplyed1} S.~Montangero, T.~Calarco, and R.~Fazio,
%``Robust Optimal Quantum Gates for Josephson Charge Qubits''
Phys. Rev. Lett. {\bf 99}, 170501 (2007).

\bibitem{QCapplyed2}  P.~Rebentrost,  {\it et. al.}, %I. Serban, T. Schulte-Herbr\"uggen,  and F. K. Wilhelm
%``Optimal Control of a Qubit Coupled to a Non-Markovian Environment''
Phys. Rev. Lett. {\bf 102}, 090401 (2009).

\bibitem{NMR} N.~Khaneja,  {\it et. al.}, %T. Reiss, C. Kehlet, T. Schulte-Herbr\"uggen and S. J. Glaser
%``Optimal control of coupled spin dynamics: design of NMR pulse sequences by gradient ascent algorithms''
J. Magn. Reson. {\bf 172}, 296 (2005).

\bibitem{kzmech} W.H.~Zurek, U.~Dorner, P.~Zoller 
%``Dynamics of a  Quantum phase transition'' 
Phys. Rev. Lett. {\bf  95} 105701 (2005).

\bibitem{farhi} E.~Farhi,  {\it et. al.}, %J. Goldstone, S. Gutmann,  J. Lapan,  A. Lundgren, D. Preda 
%``A Quantum Adiabatic Evolution Algorithm Applied to Random Instances of an NP-Complete Problem''
Science {\bf 292} 472 (2001).

\bibitem{adtheo} M.~Born and V.A.~Fock, 
%``Beweis des  Adiabatensatzes'' 
%Zeitschrift f\"ur Physik {\bf 51} 165 (1928).
Zeit. f\"ur Physik {\bf 51} 165 (1928).

\bibitem{revzurek}  W.H.~Zurek, %``Cosmological experiments in condensed matter systems''
Phys. Rep. {\bf 276}, 177 (1996).

\bibitem{vidal} G.~Vidal, 
%``Efficient Classical Simulation of Slightly Entangled Quantum
%Computations'', 
Phys. Rev. Lett. {\bf 91}, 147902 (2003).

\bibitem{revscholl} U.~Schollw\"ock, 
%``The density-matrix renormalization group''
Rev. Mod. Phys. {\bf 77}, 259 (2005).

\bibitem{MarkerAtoms} T. Calarco,  {\it et. al.}, %U. Dorner, P. S. Julienne, C. J. Williams, and P. Zoller  
%"Quantum computations with atoms in optical lattices: Marker qubits and molecular interactions"
Phys. Rev. A {\bf 70}, 012306 (2004).

\bibitem{caneva} T.~Caneva,  {\it et. al.}, %T. Calarco, R. Fazio,  G. E. Santoro, and S. Montangero, 
arXiv:1011.6634.

\bibitem{MPS} S.~\"Ostlund and S.~Rommer, Phys. Rev. Lett. {\bf 75},
  3537 (1995); J. I. Cirac and F. Verstraete, Journ. Of Phys. A  {\bf 42}, 504004 (2009).

\bibitem{numrec}
W.H.~Press,  {\it et. al.}, %S. A. Teukolsky, W. T. Vetterling, B. P. Flannery
``Numerical Recipes'',  
Cambridge University Press, New York 2007.

\bibitem{greiner} 
M.~Greiner,  {\it et. al.}, %O.~Mandel, T.~Esslinger, T. W.~H\"ansch and I. Bloch
%``Quantum phase transition from a superfluid to a Mott insulator in a gas of ultracold atoms''
Nature {\bf 415}, 39 (2002).

\bibitem{blochrev} I.~Bloch, J.~Dalibard, and W.~Zwerger, 
%``Many-body physics with ultracold gases'',
  Rev. Mod. Phys. {\bf 80}, 885 (2008).

\bibitem{toolbox} D.~Jaksch, P.~Zoller, 
%"The cold atom Hubbard toolbox", 
Annals of Physics, {\bf 315}, 52 (2005).

\bibitem{batrouni}
G.G.~Batrouni,  {\it et. al.}, %V. Rousseau, R.T. Scalettar, M. Rigol,
                             %A. Muramatsu, P.J.H. Denteneer, and M. Troyer, 
%``Mott Domains of Bosons Confined on Optical Lattices''
Phys. Rev. Lett. {\bf 89}, 117203 (2002).

\bibitem{kollath} C.~Kollath,  {\it et. al.}, %A. Iucci, T. Giamarchi, W. Hofstetter, and U. Schollw\"ock
%``Spectroscopy of Ultracold Atoms by Periodic Lattice Modulations''
Phys. Rev. Lett. {\bf 97}, 050402 (2006).

\bibitem{fertig05} C.D.~Fertig {\it et. al.},
%``Strongly Inhibited transport  of a Degenerate 1D bose Gas in a
%Lattice'', 
Phys. Rev. Lett. {\bf 94} 120403 (2005).

\bibitem{clark} I.~Danshita and C.W.~Clark,
%``Heavily Damped Motion of One-Dimensional Bose Gases in an Optical Lattice''
Phys. Rev. Lett. {\bf 102}, 030407 (2009).

\bibitem{Jaksch98} D.~Jaksch,  {\it et. al.}, %C. Bruder, J. I. Cirac, C. W. Gardiner, and P. Zoller
%``Cold Bosonic Atoms in Optical Lattices''
Phys. Rev. Lett. {\bf 81}, 3108 (1998).

\bibitem{private} M.Greiner, private communication. 

%\bibitem{montangero} S. Montangero, R. Fazio, P. Zoller, and G. Pupillo 
%``Dipole oscillations of confined lattice bosons in one dimension''
%Phys. Rev. A {\bf 79}, 041602 (2009).

\bibitem{vidal1}  
M.~Zwolak, G.~Vidal,
%``Mixed-State Dynamics in One-Dimensional Quantum Lattice Systems: 
%A Time-Dependent Superoperator Renormalization Algorithm''
Phys. Rev. Lett. {\bf 93}, 207205 (2004).

%\bibitem{white} S.R.~White, A.E.~Feigun, 
%``Real-Time Evolution Using the Density Matrix Renormalization Group''
% Phys. Rev. Lett.  {\bf 93}  076401 (2004).

%\bibitem{rowan} T. Rowan, 
%``Functional Stability Analysis of Numerical  Algorithms'', 
% Ph.D. thesis, Department of Computer Sciences, University of Texas at
%  Austin, 1990. 

\bibitem{Brif2010} C.~Brif, R.~Chakrabarti, and H.~Rabitz, New
  Journ. of Phys., {\bf 12} 075008 (2010).



\end{thebibliography}
\end{document}